\documentclass[floatfix,amssymb,prl,onecolumn,superscriptaddress,preprintnumbers,nofootinbib,aps]{revtex4-2}

\usepackage{amssymb,amsmath,mathtools,graphicx,microtype,bm,booktabs}
\usepackage[dvipsnames]{xcolor}
\definecolor{linkcolor}{rgb}{0.0,0.3,0.5}
\usepackage[unicode,colorlinks=true,linkcolor=linkcolor,citecolor=linkcolor,urlcolor=linkcolor,pdfusetitle]{hyperref}
\usepackage[T1]{fontenc}
\usepackage[utf8]{inputenc}

\newcommand{\ustc}{Department of Astronomy, University of Science and Technology of China, Hefei, Anhui 230026, China}
\newcommand{\ustcs}{School of Astronomy and Space Sciences, University of Science and Technology of China, Hefei, Anhui 230026, China}
\newcommand{\ucas}{School of Astronomy and Space Science, University of Chinese Academy of Sciences, Yuquan Road, Beijing 100049, China}
\newcommand{\ihep}{Key Laboratory for Particle Astrophysics, Institute of High Energy Physics, Chinese Academy of Sciences, Yuquan Road, Beijing 100049, China}

\begin{document}

\title{Little Red Dots as a Transient Phase of Self-Interacting Dark Matter Assisted Black Hole Growth}

\author{Yu Rong}
\email{corresponding author: rongyua@ustc.edu.cn}
\affiliation{\ustc}
\affiliation{\ustcs}
\author{Shuang-Nan Zhang}
\affiliation{\ihep}
\affiliation{\ucas}
\author{Jian-Min Wang}
\affiliation{\ihep}
\affiliation{\ucas}
\author{Junxian Wang}
\affiliation{\ustc}
\affiliation{\ustcs}
\author{Zhicheng He}
\affiliation{\ustc}
\affiliation{\ustcs}

\date{September 17, 2026}

\begin{abstract}
The discovery of Little Red Dots (LRDs) with JWST has revealed a population of compact, red galaxies hosting rapidly growing black holes at early cosmic times. Their compact morphologies, broad emission lines, weak X-ray emission, and distinctive V-shaped spectral energy distributions indicate a short-lived phase of black-hole growth within a dense nuclear environment. Here we propose that LRDs arise from a transient episode of self-interacting dark matter (SIDM)-assisted black-hole growth during galaxy assembly. In this scenario, gas inflows first establish a compact nuclear thick disk, which modifies the central SIDM distribution and provides the obscuring structure around the accreting black hole. Once the SIDM density near the seed black hole becomes sufficiently enhanced, rapid SIDM accretion drives a major increase in black-hole mass, initiating the transient LRD phase. The resulting obscured growth phase naturally suppresses direct short-wavelength emission and redistributes the radiation field, producing the red continuum and V-shaped spectral signatures of LRDs. This framework links gas inflow, SIDM dynamics, and early black-hole assembly, predicting that LRDs preferentially occur in galaxies undergoing strong nuclear inflow and evolve into ordinary AGN after this transient phase. Unlike models that require globally rare halo histories, long super-Eddington gas growth, or purely phenomenological obscuration, our model ties the overmassive black hole, X-ray weakness, spectral shape, and finite duty cycle to one local gas-triggered SIDM event.
\end{abstract}

\maketitle

\section{Introduction}
Little Red Dots (LRDs) are among the most surprising compact populations revealed by JWST.  They are selected as small, red rest-optical sources at $z\sim4$--9, often with a blue or less-reddened ultraviolet component and with number densities that are high compared with ordinary luminous quasars \citep{Matthee2024LRD,Greene2024LRD,Kocevski2024LRDLF,Akins2025LRD}.  Spectroscopy has sharpened the puzzle.  Many point-like V-shaped sources show broad permitted lines, strong Balmer-limit curvature, Balmer absorption or dense-gas signatures, and continua that cannot be described by a single standard stellar population \citep{Wang2024RUBIESLRD,Setton2024Vshape,Hviding2025RUBIESLRD,deGraaff2025TheCliff,Labbe2024UltraluminousLRD}.  The inferred black-hole masses can be large compared with the stellar or dynamical masses of their hosts \citep{Furtak2023A2744QSO1,Juodzbalis2025LRDdyn}.  At the same time, MIRI detections, weak X-ray emission, weak or absent standard hot-dust signatures in some systems, uncertain variability, and AGN-contaminated host-galaxy masses complicate a simple unobscured-quasar interpretation \citep{PerezGonzalez2024MIRILRD,PacucciNarayan2024XrayWeak,Li2024DustyFlows,deGraaff2025TheCliff}.  Thus the emerging observational consensus is not that LRDs are one familiar object seen at high redshift, but that they are compact, common, frequently AGN-like, and embedded in dense reprocessing gas.

This combination of facts has led to several classes of models.  Obscured or dust-reddened AGN models provide the most direct explanation for broad permitted lines and high inferred black-hole masses \citep{Greene2024LRD,Li2024DustyFlows,PacucciNarayan2024XrayWeak}.  In this spirit, \citet{MadauMaiolino2026LRDLBD} proposed a super-Eddington orientation model in which LRDs are obscured views of little blue dots, with a thick accretion flow, a polar funnel, an equatorial broad-line region, and a dusty screen.  Such models explain the LRD/LBD connection and orientation dependence elegantly, but they still put much of the burden on gas accretion: the gas must grow an overmassive black hole, suppress escaping X-rays, generate a red continuum without always producing an ordinary hot-dust torus, and remain in place despite radiative feedback.  Dense-gas or ``black-hole star'' models instead focus on the Balmer-limit curvature and absorption, interpreting these features as dense gas around a central ionizing source rather than as an old stellar break \citep{InayoshiMaiolino2024DenseGas,deGraaff2025TheCliff}.  This insight is important, but it leaves a dynamical question: what creates the dense gas, why is it maintained near the black hole, and why should the phenomenon be short-lived and concentrated at high redshift?  Compact stellar, low-angular-momentum, low-spin halo, and forming-globular-cluster scenarios emphasize the small sizes, high surface densities, and possible stellar contribution of LRDs \citep{Williams2024CompactLRD,Baggen2024LRDDensities,PacucciLoeb2025LowSpin,Chisholm2026GlobularLRD}.  These models capture the compactness, but stellar compactness by itself does not naturally explain broad lines, weak X-rays, large black-hole-to-host ratios, and a transient AGN-like phase.

A second family of explanations places the origin of LRDs in special seeds or special halos.  Direct-collapse black-hole and quasi-star scenarios connect LRDs to rapid early seed formation, dense inflow, and buried accretion \citep{Mayer2014DirectCollapse,Mayer2023CosmoDCBH,Jeon2025DCBHLRD,BegelmanDexter2025QuasiStars,PacucciFerraraKocevski2026DCBH}.  Their strength is that they naturally contain compact, optically thick gas; their weakness is that the required conditions are restrictive, involving rapid inflow, suppressed fragmentation, unusual thermodynamic histories, and often a limited host-halo population.  Primordial-black-hole, collapsed-dark-star, and related exotic channels can place massive compact objects at very early times \citep{Carr2016PBH,ZhangFengAn2026PBHClusters,Ilie2026DarkStars}, but they still must explain the observed LRD spectra, X-ray weakness, Balmer curvature, and duty cycle.  Low-spin compact-galaxy models face a similar demographic issue: low angular momentum can make compact systems, but it also selects a special subset of halos \citep{PacucciLoeb2025LowSpin}.  Recent clustering and assembly-bias studies make this point quantitative.  Several channels, including direct collapse, SIDM core collapse, low-spin compact galaxies, and primordial-black-hole-related models, can predict distinctive or strong clustering because they select rare halo histories \citep{CarranzaEscudero2025LRDclustering,Wang2026LRDAssemblyBias,Zhang2026LRDhalo}.  Current data do not yet require all LRDs to live in the rarest large-scale peaks.  A successful model should therefore explain the spectra and black-hole masses without automatically making every LRD a globally extreme halo.

Self-interacting dark matter (SIDM) provides a way to address the black-hole mass problem through a dark, weakly radiating growth channel.  SIDM was originally proposed to alleviate small-scale structure tensions in collisionless cold dark matter \citep{SpergelSteinhardt2000SIDM}; velocity-dependent SIDM can modify dwarf and galaxy-scale halos while satisfying larger-scale constraints \citep{Kaplinghat2016SIDM,TulinYu2018SIDM}.  Baryons can strongly reshape SIDM centers: compact baryonic potentials can contract SIDM and generate dense cusps, whereas feedback-dominated systems can form lower-density cores \citep{Sameie2018SIDMBaryons,Despali2019SIDMBaryons}.  SIDM has been used to interpret the diversity of galaxy rotation curves and the clustering of diffuse dwarfs \citep{Kaplinghat2016SIDM,TulinYu2018SIDM,Zhang2025DiffuseDwarfsSIDM}.  For LRDs, \citet{Jiang2026SIDMLRD} proposed that gravothermal collapse of SIDM halos can directly form massive black holes, elegantly avoiding a long phase of luminous super-Eddington gas accretion.

The pure SIDM gravothermal-collapse channel is attractive, but it also illustrates the problem we want to avoid.  The collapse time is highly sensitive to halo concentration, formation time, and assembly history; the very conditions that make early collapse possible can imply strong assembly bias and special environments \citep{Wang2026LRDAssemblyBias}.  It also leaves a redshift question.  If the visible LRD phase simply traced halos that had enough time to collapse gravothermally, lower-redshift halos should have had more time to produce analogous systems, whereas observed LRD demographics appear to decline rapidly toward later times \citep{Kocevski2024LRDLF,Akins2025LRD,Zhang2026LRDhalo}.  These issues do not rule out SIDM.  Instead, they suggest that SIDM should be coupled to baryonic compaction: SIDM may provide the dark mass reservoir, while gas inflow, obscuration, and viewing geometry determine when the system is visible as an LRD.

Here we develop this coupled picture by revisiting black-hole growth through accretion of collisional dark matter.  \citet{Hu2006SIDMacc} showed that a compact seed embedded in collisional dark matter can grow rapidly and saturate when the SIDM mean free path becomes comparable to the accretion radius; related calculations have been extended by \citet{DasKalita2024SIDM} and \citet{Sabarish2025SIDMspikes}.  In our model the black hole is not born from collapse of the whole SIDM halo.  Instead, a pre-existing central black hole must be large enough that, after baryonic compression, the SIDM density inside its influence radius satisfies the mean-free-path ignition condition.  The seed mass therefore has a lower bound set by the finite compressive power of the gas disk.  Once this condition is met, the subsequent saturated black-hole mass is set mainly by the SIDM sound speed and self-interaction cross-section.  A gas-rich merger or compaction event forms a compact, turbulent, optically thick nuclear disk like those found in merger-driven direct-collapse simulations \citep{Mayer2014DirectCollapse,Mayer2023CosmoDCBH}.  This disk deepens the central baryonic potential, compresses the local SIDM, and triggers rapid SIDM-fed black-hole growth.  The black-hole mass jump then further deepens the potential and helps maintain gas pile-up and high column density around the nucleus.  In this way SIDM supplies the dark mass growth, while the nuclear thick disk supplies the obscuration, Balmer-active reprocessing layer, X-ray weakness, orientation dependence, and finite lifetime.

The goal of this paper is to construct a semi-analytic version of this gas-triggered SIDM-accretion model.  We treat the obscurer as a compact nuclear thick disk rather than as a separate spherical cocoon, derive its column density from the gas mass and disk size, compute the SIDM ignition and saturation conditions, and generate absolute-flux mock spectra.  We compare the resulting spectra with public LRD spectroscopy from \citet{Wang2024RUBIESLRD} and \citet{deGraaff2025LRDPopulation}.  The chronology is central: baryonic compaction turns on SIDM accretion; SIDM accretion creates the overmassive black hole; the same compact disk hides and reprocesses the subsequent AGN radiation; and the LRD phase ends when the high-covering, Balmer-active disk atmosphere is consumed, diluted, or expelled.
Throughout this paper, we use $G$ for Newton's constant, $m_{\rm p}$ for the proton mass, $\sigma_{\rm T}$ for the Thomson cross-section.

\section{Semi-analytic model for the formation of an LRD}

\subsection{Host Halo and Disk Compactness}

We use RUBIES-EGS-55604 \citep{Wang2024RUBIESLRD} as the fiducial case for developing the model. We adopt a SIDM halo with mass
\begin{equation}
M_{\rm h}\simeq 5.0\times10^{10}M_\odot
\end{equation}
at $z=6.982$, the spectroscopic redshift of RUBIES-EGS-55604 \citep{Wang2024RUBIESLRD}.
This halo mass is compatible with a massive but not uniquely rare high-redshift host and with AGN-contaminated stellar mass estimates \citep{Behroozi2019UniverseMachine,PacucciNarayan2024XrayWeak}.  The high-redshift virial scaling of \citet{BarkanaLoeb2001} gives a virial velocity $V_{\rm vir}$ and  effective one-dimensional SIDM sound speed $C_{\rm s}$ of
\begin{equation}
V_{\rm vir}=120{\rm km\,s^{-1}},\qquad
C_{\rm s}=\frac{V_{\rm vir}}{\sqrt2}=85{\rm km\,s^{-1}} .
\end{equation}
The baryon mass in this halo is approximately $M_{\rm B}=f_{\rm B}M_{\rm h}=7.8\times 10^{9}M_{\odot}$, where $f_{\rm B}=0.156$ \citep{Planck2020}. There is a small central seed BH with typical mass of $M_{\rm BH,seed}\simeq 10^5M_{\odot}$ in this halo, along with the initial stellar mass of approximately $M_{\star}\sim 10^7\--10^9M_{\odot}$ with $M_{\rm BH,seed}/M_{\star}\sim 10^{-4}\--10^{-2}$.

Motivated by the simulations of \citet{Mayer2014DirectCollapse,Mayer2023CosmoDCBH}, a merger with this halo at high-redshift can trigger a gas inflow, forming a compact nuclear thick disk in the halo center. We adopt the thick-disk gas mass of
\begin{equation}
M_{\rm gas,d}=10^8M_\odot,
\end{equation}
approximately $1\%$ of the baryonic budget.
The compact disk radius is then inferred from the characteristic internal circular speed $v_{\rm c,d}$ of such nuclear disks,
\begin{equation}
v_{\rm c,d}=\chi C_{\rm s},
\end{equation}
where the compactness parameter $\chi$ characterizes how deep the local baryonic disk potential is relative to the SIDM velocity scale.  The compact nuclear disks in the simulations of \citet{Mayer2014DirectCollapse,Mayer2023CosmoDCBH} can be substantially more compact than our fiducial choice when their quoted disk masses and radii are translated into $v_{\rm c,d}$ and hence $\chi$.  We deliberately adopt a conservative sub-extreme value, $\chi\simeq1$--3, so that the fiducial calculation does not rely on the most compact simulated disks.  A larger $\chi$ corresponds to a deeper local baryonic potential and, for fixed disk mass, a smaller characteristic disk radius and larger surface density.  Such a state is expected to have a larger column density and a higher ionization/heating rate per unit area, and may correspond to an earlier, more compact phase of the nuclear thick disk.  As the disk evolves, gas consumption, accretion, outflows, and turbulent expansion can lower the surface density and effective optical depth, while the escaping dense-gas continuum can shift from a hotter photoionized color temperature toward the cooler $3000$--$6000$ K range seen in compact-disk simulations \citet{Mayer2014DirectCollapse,Mayer2023CosmoDCBH}.  The object ceases to appear as an LRD once the disk no longer maintains the high-column, Balmer-active reprocessing layer.  We estimate this column-clearing timescale in Section~2.4.

For RUBIES-EGS-55604, we adopt $\chi\sim 2$. The characteristic radius of the nuclear thick disk is then
\begin{equation}
R_{\rm d}=\frac{GM_{\rm gas,d}}{v_{\rm c,d}^2}\simeq 14\ \rm {pc}.
\label{eq:td_rd}
\end{equation}

For such a turbulent or rotation-supported thick disk in vertical dynamical equilibrium, the scale height satisfies
\begin{equation}
\frac{H}{R_{\rm d}}\simeq\frac{\sigma_{\rm g}}{v_{\rm c,d}},
\label{eq:td_hr}
\end{equation}
where $H$ is the vertical scale height, and $\sigma_{\rm g}$ is the gas turbulent or vertical velocity dispersion \citep{BinneyTremaine2008}.  We adopt $\sigma_g=100{\rm km\,s^{-1}}$. Thus it gives
\begin{equation}
\frac{H}{R_{\rm d}}\sim 0.6,\qquad H=8.0{\rm pc}.
\end{equation}
The disk surface density is
\begin{equation}
\Sigma_{\rm g}=\frac{M_{\rm gas,d}}{\pi R_{\rm d}^2}
\simeq 1.6\times10^5\ M_\odot{\rm pc^{-2}} .
\end{equation}
The vertical hydrogen-equivalent column is
\begin{equation}
N_{\rm H,\perp}=\frac{\Sigma_{\rm g}}{1.4m_{\rm p}}
\simeq 1.5\times10^{25}{\rm cm^{-2}},
\label{eq:td_nhperp}
\end{equation}
where the factor $1.4m_{\rm p}$ approximately includes helium.  The corresponding Thomson depth is $\tau_{\rm T,\perp}=N_{\rm H,\perp}\sigma_{\rm T}\sim 9.6$, so even the vertical column is Compton thick \citep{OsterbrockFerland2006}.  A line of sight through the disk body has a larger characteristic column,
\begin{equation}
N_{\rm H,edge}\simeq \frac{N_{\rm H,\perp}}{H/R_{\rm d}}
\sim 2.5\times 10^{25}{\rm cm^{-2}} ,
\label{eq:td_nhedge}
\end{equation}
with the corresponding Thomson depth of $\tau_{\rm T,edge}=N_{\rm H,edge}\sigma_{\rm T}\sim 16.6$.
The midplane density is
\begin{equation}
\rho_{\rm mid}\simeq\frac{M_{\rm gas,d}}{2\pi R_{\rm d}^2H}
\simeq 1.0\times10^4\ M_\odot{\rm pc^{-3}}.
\end{equation}
These values imply that the nuclear thick-disk is an obscurer with large optical depth.  The highest columns occur through the disk midplane and the inner rim, while a turbulent disk atmosphere or disk-wind base can add high-covering, lower-temperature material above the midplane.  The vertical column is already Compton thick, but a face-on observer can still receive a much bluer spectrum if the polar direction has a lower effective covering factor than the disk body.

\subsection{SIDM Compression triggers BH rapid growth}

The thick disk can trigger SIDM compression.  Baryonic contraction of dark matter is a standard response to condensation of baryons \citep{Blumenthal1986AdiabaticContraction,Gnedin2004AdiabaticContraction}, and SIDM simulations with baryons show that central baryonic potentials can make SIDM profiles denser or cuspier than dark-matter-only SIDM halos \citep{Sameie2018SIDMBaryons,Despali2019SIDMBaryons}.

For approximately isothermal SIDM in hydrostatic balance,
\begin{equation}
\frac{d(\rho_\chi C_{\rm s}^2)}{dr}
=-\rho_\chi\frac{d\Phi}{dr},
\end{equation}
where $\rho_\chi$ is SIDM density and $\Phi$ is gravitational potential.  The density amplification from $R_{\rm d}$ to the black-hole influence region is
\begin{equation}
{\cal C}_\chi\equiv
\frac{\rho_\chi(r_{\rm B})}{\rho_\chi(R_{\rm d})}
=\exp\left(\frac{\Delta\Phi}{C_{\rm s}^2}\right)
=\exp(\eta_\Phi\chi^2),
\label{eq:td_cchi}
\end{equation}
where $r_{\rm B}=GM_{\rm BH}/C_{\rm s}^2$ is the black-hole influence radius and $\eta_\Phi$ parametrizes the disk potential shape. $\rho_\chi(r_{\rm B})$ and $\rho_\chi(R_{\rm d})$ denote the SIDM density at $r_{\rm B}$ and $R_{\rm d}$, respectively. A uniform spherical potential would give $\eta_\Phi=1/2$, while a centrally concentrated disk or cusp gives a larger potential drop and motivates $\eta_\Phi\sim1$ as an upper illustrative value \citep{BinneyTremaine2008}.  For $\chi\sim 2$, Equation~\ref{eq:td_cchi} gives ${\cal C}_\chi\simeq 70$, suggesting that the SIDM density in the central BH influence radius is approximately amplified 70 times by the nuclear thick-disk.

The SIDM mean-free-path ignition condition is
\begin{equation}
\lambda_{\rm mfp}(r_{\rm B})
=\frac{1}{\rho_\chi(r_{\rm B})\sigma/m}
\lesssim r_{\rm B} ,
\label{eq:td_ignition}
\end{equation}
where $\sigma/m$ is the SIDM self-interaction cross-section per unit mass \citep{Hu2006SIDMacc,DasKalita2024SIDM,Sabarish2025SIDMspikes}. If equation~(\ref{eq:td_ignition}) is satisfied, the central BH is able to efficiently accrete SIDM. For $\sigma/m\sim 1\ {\rm cm^2\,g^{-1}}$ \citep{Kaplinghat2016SIDM,TulinYu2018SIDM,Zhang2025DiffuseDwarfsSIDM} and $C_{\rm s}\sim 85{\rm km\,s^{-1}}$, the density required around a $10^5M_\odot$ black hole is
\begin{equation}
\rho_{\rm ign}\simeq 8\times 10^4M_\odot{\rm pc^{-3}}
\left(\frac{\sigma/m}{1{\rm cm^2\,g^{-1}}}\right)^{-1}
\left(\frac{C_{\rm s}}{85{\rm km\,s^{-1}}}\right)^2
\left(\frac{M_{\rm BH}}{10^5M_\odot}\right)^{-1}.
\label{rho_ign}
\end{equation}
Thus the pre-compression density needed at $R_{\rm d}$ is only
\begin{equation}
\rho_\chi(R_{\rm d})=\frac{\rho_{\rm ign}}{{\cal C}_\chi}
\simeq 1.2\times10^3M_\odot{\rm pc^{-3}},
\label{rho_sidm_ign}
\end{equation}
or $12\%$ of $\rho_{\rm mid}$, implying that the nuclear thick-disk can compress the inner SIDM density, and triggers central BH accreting SIDM.

Equations~(\ref{rho_ign}) and~(\ref{rho_sidm_ign}) also imply a lower limit on the BH seed mass.  If the required pre-compression density satisfies $\rho_\chi(R_{\rm d})\gg\rho_{\rm mid}$, the nuclear thick disk would not be able to compress the inner SIDM density efficiently enough to ignite SIDM accretion.  Therefore, the lower limit of the BH seed mass can be estimated by requiring $\rho_\chi(R_{\rm d})\simeq\rho_{\rm mid}$, and this limit strongly depends on the density of the nuclear thick disk.  For a fiducial nuclear thick disk with $M_{\rm gas,d}\sim10^8M_{\odot}$ and $R_{\rm d}\sim10$ pc, the lower limit of the BH seed mass is approximately $10^4M_{\odot}$.  For the most extreme nuclear thick disks in the simulations of \citet{Mayer2014DirectCollapse,Mayer2023CosmoDCBH}, with $M_{\rm gas,d}\sim10^9M_{\odot}$ and $R_{\rm d}\sim5$ pc, the lower limit could decrease to $\sim100M_{\odot}$, comparable to a Pop~III remnant black hole \citep{MadauRees2001PopIII,Heger2003PopIIIRemnants,BrommLarson2004FirstStars}.

For collisionless cold dark matter, black-hole growth by dark-matter capture is inefficient because the dark component lacks dissipation and only particles entering the loss cone can be captured. SIDM changes this because self-interactions can make the dark component behave as an effective collisional fluid on relevant scales. 

\begin{figure}
\centering
\includegraphics[width=0.92\textwidth]{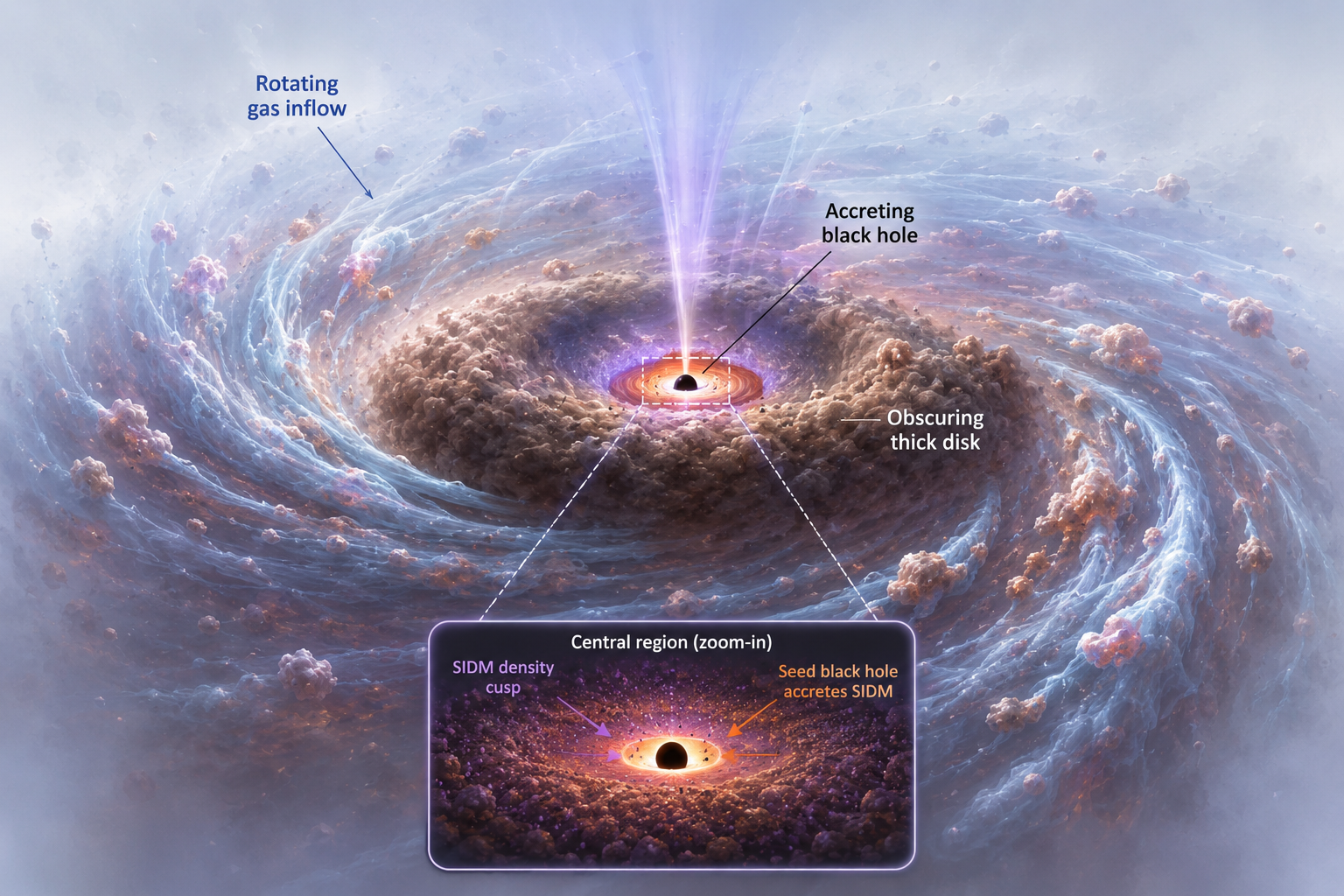}
\caption{Schematic picture of the model.  A gas-rich inflow builds a compact optically thick nuclear disk.  The disk potential compresses SIDM and triggers rapid SIDM-fed black-hole growth.  Edge-on sight lines pass through high-column disk gas and appear as red, X-ray-weak LRDs.}
\label{fig:td_cartoon}
\end{figure}

This SIDM accretion process is very fast. The rapid SIDM accretion stage saturates when the mean free path becomes comparable to the accretion radius.  \citet{Hu2006SIDMacc} give the BH mass after accretion 
\begin{equation}
M_{\rm BH,SIDM}
=1.4\times10^6M_\odot
\left[\frac{\sigma/m}{1{\rm cm^2\,g^{-1}}}\right]
\left(\frac{C_{\rm s}}{30{\rm km\,s^{-1}}}\right)^4 ,
\label{eq:td_msat}
\end{equation}
with characteristic time
\begin{equation}
t_{\rm SIDM}=\frac{(\sigma/m)C_{\rm s}}{4\pi G}.
\label{eq:td_tsidm}
\end{equation}
For $\sigma/m\sim 1\ {\rm cm^2\,g^{-1}}$ and $C_{\rm s}\sim 85\ {\rm km\,s^{-1}}$,  equations~\ref{eq:td_msat}--\ref{eq:td_tsidm} give
\begin{equation}
M_{\rm BH,SIDM}\simeq 9\times 10^7M_\odot,\qquad
t_{\rm SIDM}\simeq 0.32{\rm Myr}.
\end{equation}

Since the timescale of BH accretion is too fast, the halo and stellar components almost do not evolve during this phase. 

SIDM accretion is electromagnetically dark. During the SIDM accretion period, the central BH should also accretes the surrounding gas. However, the accretion rate is limited by Eddington accretion rate and is much lower than the rate of accreting SIDM, and thus can be neglected. 

\subsection{The LRD mode}

{\bf In our model, the LRD mode begins only when three conditions are simultaneously satisfied: the black hole has reached the rapid-SIDM saturation mass, baryonic gas is still present in an optically thick compact disk, and an edge-on or high-covering line of sight intersects the disk body or atmosphere.}  This makes the onset a potential-transformation event rather than simply the presence of a gas disk or the presence of an AGN, as shown in Fig.~\ref{fig:td_cartoon}.  The process is also a feedback loop.  The inflow-built disk first compresses SIDM and initiates rapid dark accretion.  The resulting black-hole mass jump then deepens the central potential on a timescale shorter than the disk replenishment or clearing time, which promotes further gas inflow, raises the central column, and helps maintain the obscuring disk atmosphere during the luminous phase.

After SIDM saturation, the black-hole accreting gas becomes more dominant, which begins to emit as an AGN. The surrounding nuclear thick-disk is retained because the disk remains self-gravitating during the luminous phase.  The buried AGN luminosity is then set by baryonic accretion onto the SIDM-enlarged black hole,
\begin{equation}
L_{\rm bol}=\lambda_{\rm eff}L_{\rm Edd}
=1.26\times10^{38}\lambda_{\rm eff}
\left(\frac{M_{\rm BH,SIDM}}{M_\odot}\right)
{\rm erg\,s^{-1}},
\label{eq:td_lbol}
\end{equation}
where $\lambda_{\rm eff}$ is the effective Eddington ratio and $L_{\rm Edd}$ is the Eddington luminosity \citep{Salpeter1964,Volonteri2010SMBH}.  The RUBIES-EGS-55604 continuum fit prefers a luminous but sub-Eddington buried AGN, $\lambda_{\rm eff}=0.6$,
\begin{equation}
L_{\rm bol}=6.8\times10^{45}{\rm erg\,s^{-1}},
\qquad
\dot M_{\rm BH}=\frac{L_{\rm bol}}{\epsilon c^2}
=1.2M_\odot{\rm yr^{-1}}
\left(\frac{\epsilon}{0.1}\right)^{-1},
\label{eq:td_mdotbh}
\end{equation}
where $\dot M_{\rm BH}$ is the baryonic black-hole accretion rate, $\epsilon$ is the radiative efficiency, and $c$ is the speed of light. Our model does not require a strongly super-Eddington radiative accretion ratio, $\lambda_{\rm eff}\gg 1$, which remains observationally uncertain in high-redshift LRDs.

The disk columns in Equations~\ref{eq:td_nhperp} and~\ref{eq:td_nhedge} imply Thomson depths $\tau_{\rm T}\gg1$, so the directly escaping blue AGN continuum is strongly suppressed along high-covering sight lines \citep{OsterbrockFerland2006}.  Multiple electron scatterings, bound-free absorption, free-free absorption, metal opacity, and dust or dust-free thermalization channels can therefore reprocess a large fraction of the AGN power into a red continuum plus an obscured photosphere from the compact disk \citep{OsterbrockFerland2006,Ferland2017Cloudy}.  We write the latter as
\begin{equation}
F_\nu^{\rm ph}=
A_{\rm ph}\,
B_\nu(T_{\rm gas})
\left(\frac{\lambda_0}{\lambda}\right)^{\beta_{\rm ph}}
{\cal T}_{\rm B}(\lambda),
\label{eq:td_phshape}
\end{equation}
where $F_\nu^{\rm ph}$ is the obscured-photosphere component, $B_\nu$ is the Planck function, $T_{\rm gas}$ is the color or thermal-equilibrium temperature of the optically thick dense gas, $\lambda_0$ is a reference wavelength, $\beta_{\rm ph}$ is an effective wavelength-dependent escape or opacity index, and ${\cal T}_{\rm B}$ is the transmission through the Balmer-active layer of the disk atmosphere or wind base.  The normalization $A_{\rm ph}$ is fixed by the reprocessed luminosity budget,
\begin{equation}
\int L_\nu^{\rm ph}\,d\nu
=L_{\rm ph}
=f_{\rm ph}L_{\rm bol},
\label{eq:td_lph}
\end{equation}
where $f_{\rm ph}$ is the fraction of the buried AGN luminosity emerging through the obscured photosphere. Equation~\ref{eq:td_phshape} is not a separate object from the gas disk; it is a compact spectral representation of unresolved radiative transfer through the optically thick disk atmosphere.  Different LRDs need not have the same $T_{\rm gas}$ because dense gas has several plausible thermal states.  Strong photoionization and efficient Ly$\alpha$/recombination cooling drive photoionized gas toward a $T_{\rm gas}\sim10^4$ K plateau, while lower ionization parameter, stronger shielding, molecular or metal cooling, and a small amount of dust can shift the escaping continuum to a cooler $T_{\rm gas}\sim3000$--$5000$ K color temperature \citep{OsterbrockFerland2006,Ferland2017Cloudy}.  In this paper $T_{\rm gas}$ is therefore an effective continuum-transfer parameter fitted together with the obscured AGN component.

Because the Balmer limit is not produced by a single cold slab, we represent the outer absorbing layer with a smooth effective transmission:
\begin{equation}
{\cal T}_{\rm B}(\lambda)=
(1-C_{\rm B})+C_{\rm B}\exp[-\tau_{\rm B} S_{\rm B}(\lambda)],
\label{eq:smooth_Balmer_trans}
\end{equation}
with
\begin{equation}
S_{\rm B}(\lambda)=
\frac{1}{1+\exp[(\lambda-\lambda_{\rm B})/\Delta\lambda_{\rm B}]}\times\begin{cases}
(\lambda/\lambda_{\rm B})^3, & \lambda<\lambda_{\rm B},\\
1, & \lambda\ge \lambda_{\rm B},
\end{cases}
\end{equation}
where $\lambda_{\rm B}=0.3646\,\mu{\rm m}$ is the Balmer edge, $\Delta\lambda_{\rm B}$ is the effective smoothing width, and $C_{\rm B}$ is the covering factor of the Balmer-active layer.  Physically, this smooth form represents a stratified disk atmosphere or disk-wind base with velocity, density, temperature, and ionization gradients.  In Fig.~\ref{fig:td_spectra}, the fiducial RUBIES-EGS-55604 solution uses the smooth Balmer transmission with $C_{\rm B}=0.70$, $\tau_{\rm B}=15$, $\Delta\lambda_{\rm B}=0.006\,\mu{\rm m}$, and no extra trough term.  The fit to another typical LRD, RUBIES-UDS-40579, uses the same functional form with a cooler color temperature and a broader effective Balmer layer; this is a change in the radiative-transfer state of the same thick-disk atmosphere, not a change in the underlying model.

We model the LRD continuum with four observable components:
\begin{equation}
F_\nu =
F_\nu^{\rm obscured\ AGN}
+
F_\nu^{\rm obscured\ photosphere}
+
F_\nu^{\rm leaked\ young\ stars}
+
F_\nu^{\rm old\ host}.
\label{eq:td_twocomp}
\end{equation}
The first term is radiation from the accreting black hole after propagation through the optically thick disk.  The second term is the obscured photosphere defined in Equation~\ref{eq:td_phshape}.  The third term represents a small fraction of recently formed young nuclear starlight that leaks or scatters through low-opacity channels.  The fourth term represents weak pre-existing host starlight in the spectroscopic aperture.  This is the moment when the model becomes an LRD: the black hole is already overmassive, the gas is still optically thick, and the observed continuum is dominated by reprocessed rather than direct AGN light.

\subsection{The End of the LRD Mode}

The LRD lifetime is the time over which the optically thick disk body and atmosphere remain replenished faster than they are consumed, accreted, expelled, or geometrically diluted.  We write this column-clearing time as
\begin{equation}
t_{\rm LRD}\simeq
\frac{M_{\rm gas,d}-M_{\rm gas,crit}}
{\dot M_{\rm BH}+\dot M_\star+\dot M_{\rm out}-\dot M_{\rm in}},
\label{eq:td_tlrd}
\end{equation}
where $\dot M_\star$, $\dot M_{\rm out}$, and $\dot M_{\rm in}$ are the compact star-formation, outflow, and replenishing inflow rates, respectively.  The critical mass $M_{\rm gas,crit}$ is the disk mass below which the object no longer looks like an LRD.  If the disk radius is roughly fixed, the vertical column evolves as
\begin{equation}
N_{\rm H,\perp}(t)\simeq
N_{\rm H,\perp,0}
\left[\frac{M_{\rm gas,d}(t)}{10^8M_\odot}\right]
\left(\frac{R_{\rm d}}{14{\rm pc}}\right)^{-2}.
\label{eq:td_nhevol}
\end{equation}
The Compton-thick condition alone is lost only after the gas mass drops below $\simeq7\times10^6M_\odot$ for $N_{\rm H,\perp}=10^{24}{\rm cm^{-2}}$, but the V-shaped LRD continuum is more sensitive to the Balmer-edge layer.  For the fiducial Balmer optical depth used in Section~3, reducing the effective column by a factor of a few is enough to weaken the Balmer-limit curvature.  Taking $M_{\rm gas,crit}\simeq(0.2$--$0.3)M_{\rm gas,d}$ gives
\begin{equation}
t_{\rm LRD}\sim 10\text{--}100{\rm Myr}
\left(\frac{M_{\rm gas,d}}{10^8M_\odot}\right)
\left(\frac{\dot M_{\rm net}}{1\text{--}10M_\odot{\rm yr^{-1}}}\right)^{-1},
\label{eq:td_tlrd_num}
\end{equation}
where $\dot M_{\rm net}\equiv\dot M_{\rm BH}+\dot M_\star+\dot M_{\rm out}-\dot M_{\rm in}$.  For RUBIES-EGS-55604, without replenishment, Eddington black-hole accretion alone would remove $10^8M_\odot$ in $\simeq16$ Myr, and star formation or outflows would shorten this time.  A long LRD lifetime therefore requires continued supply from the larger nuclear disk or repeated compaction, consistent with gas-rich merger simulations that assemble compact disks rapidly but do not imply a static Gyr-lived obscurer \citep{Mayer2014DirectCollapse,Mayer2023CosmoDCBH}.  The abundance-based lifetime estimates for LRD populations are still uncertain, but duty-cycle arguments generally point to a short phase rather than a Hubble-time population \citep{Kocevski2024LRDLF,Akins2025LRD,Hviding2025RUBIESLRD}.

The LRD mode ends when the high-covering column falls, the polar channel widens, or the disk expands so that the photospheric and Balmer-edge conditions are no longer satisfied.  The descendant should then be a less-obscured compact AGN, a blue compact broad-line AGN if viewed through the polar direction, or a compact post-starburst/overmassive-black-hole system with weaker Balmer-edge curvature.  In this picture, low-redshift analogues are uncommon because later small-galaxy mergers are less gas rich, have more feedback-regulated central gas reservoirs, and occur after the first large fractional black-hole mass jump has already passed.


\section{Mock Spectra and Observed LRD Spectra}
\label{sec:td_spectra}

\subsection{Mock-Spectrum Construction}

All spectra in this section are generated in absolute flux units from the same component library.  The obscured-AGN continuum is an analytic red reprocessing template, $F_\nu^{\rm red}\propto [1+\exp[-(\lambda-\lambda_{\rm B})/w]]^{-1}(\lambda/0.55\,\mu{\rm m})^{\alpha_{\rm red}}$, optionally multiplied by the finite red-side self-absorption factor in Equation~\ref{eq:td_agn_turnover}.  The obscured photosphere is the modified Planck template in Equation~\ref{eq:td_phshape}.  Both components are normalized by integrating the rest-frame $L_\nu$ template over frequency and setting the luminosity to the fitted fraction of $L_{\rm bol}$. The Balmer curvature is then applied with the same smooth transmission function, Equation~\ref{eq:smooth_Balmer_trans}, for both fitted LRDs.

The stellar terms are deliberately continuum-only nuisance templates.  For the leaked young component we use an FSPS-motivated young stellar-population shape with nebular emission switched off, normalized by the standard rest-UV continuous-star-formation conversion \(L_\nu(1500\,\mathrm{A})\simeq8\times10^{27}(\mathrm{SFR}/M_\odot\,\mathrm{yr}^{-1})\,\mathrm{erg\,s^{-1}\,Hz^{-1}}\) and attenuated, when needed, with a Calzetti-law screen \citep{Conroy2009FSPS,ConroyGunn2010FSPS,KennicuttEvans2012,Calzetti2000}.  The weak old-host term uses the same stellar-population-motivated continuum shape and a simple rest-optical mass-to-light normalization.  These stellar templates are included only to test whether aperture-level galaxy light can smooth the continuum; they contain no emission lines and are not allowed to dominate the LRD continuum.

We do not fit the emission lines in this paper.  The grey masks in Figures~\ref{fig:td_spectra} and~\ref{fig:td_40579} remove line complexes before computing continuum residuals.  A full line and bound-free continuum calculation should be done with a photoionization/radiative-transfer code such as CLOUDY \citep{OsterbrockFerland2006,Ferland2017Cloudy}.  Here CLOUDY is used only as the physical motivation for the dense-gas opacity discussion, not as an input library for the plotted mock spectra.

\subsection{A Hot Dense-Gas LRD: RUBIES-EGS-55604}

The thick disk is Compton thick in both its vertical and disk-body columns, but a polar funnel or low-covering channel can still allow a larger fraction of blue radiation to escape along the disk axis.  For the RUBIES-EGS-55604 continuum we fit the high-column spectrum using the components in Equation~\ref{eq:td_twocomp}.  The dynamical and SIDM quantities are tied to the equations in Section~2.  By contrast, the continuum-transfer quantities are effective parameters: the red reprocessed slope $\alpha_{\rm red}$, the red AGN fraction $f_{\rm red}$, the obscured-photosphere fraction $f_{\rm ph}$, the Balmer-layer covering factor $C_{\rm B}$, and the Balmer-edge width depend on multidimensional radiative transfer through the disk atmosphere, viewing angle, scattering, thermalization depth, ionization structure, and possible dust or dust-free opacity.  We therefore fit them to the line-masked continuum, while requiring subdominant stellar contributions and physically plausible columns.

This separation is important.  The semi-analytic model determines the gravitational and energetic scale of the event: $M_{\rm h}$ sets $C_{\rm s}$, the compact disk fixes $R_{\rm d}$ and $N_{\rm H}$, the adopted galaxy-scale $\sigma/m$ fixes the SIDM saturation mass through Equation~\ref{eq:td_msat}, and $\lambda_{\rm eff}$ fixes the bolometric luminosity through Equation~\ref{eq:td_lbol}.  What it does not determine is how a clumpy, optically thick, non-LTE disk atmosphere partitions that luminosity among a red scattered or reprocessed AGN continuum, a compact thermalization surface, Balmer-edge absorption, and small leakage channels.  Those quantities require radiation-hydrodynamic transfer or photoionization calculations \citep{OsterbrockFerland2006,Ferland2017Cloudy}.  In this paper they are therefore fitted effective parameters, not hidden dynamical parameters.

For RUBIES-EGS-55604 the best continuum solution has
\begin{equation}
\alpha_{\rm red}=4.3,\quad
T_{\rm gas}=1.4\times10^4{\rm K},\quad
f_{\rm red}=0.0245,\quad
f_{\rm ph}=0.0455,\quad
C_{\rm B}=0.70,
\end{equation}
with $\tau_{\rm B}=15$ and an effective Balmer-edge width $0.006\,\mu{\rm m}$.  The fit uses a modest leaked young component, $\mathrm{SFR}_{\rm eff}=6.9M_\odot{\rm yr^{-1}}$, age $10$ Myr, and leakage fraction $0.2$, plus a weak old host component, $M_{\star,\rm old}=7.6\times10^8M_\odot$, contributing only a few percent of the continuum near $5500\,\text{\AA}$.  These stellar terms are included because the observed spectrum is an aperture spectrum of the whole compact source; they are not allowed to dominate the LRD continuum.

The fitted Balmer depth is tied back to the disk columns derived above.  Equations~\ref{eq:td_nhperp} and~\ref{eq:td_nhedge} give the hydrogen column available in the compact disk, so the effective column seen by an LRD sight line should lie between the vertical and disk-body values, depending on orientation and covering.  If a fraction of that column lies in a cooler Balmer-active atmosphere or disk-wind base, the Balmer-edge optical depth is
\begin{equation}
\tau_{\rm B}\simeq N_{\rm H,B}\left(\frac{N_2}{N_{\rm H}}\right)\sigma_{\rm B},
\label{eq:td_taub}
\end{equation}
where $N_{\rm H,B}$ is the effective hydrogen column in that Balmer-active layer, $N_2/N_{\rm H}$ is the fraction of hydrogen in the first excited state, and $\sigma_{\rm B}\simeq2\times10^{-19}{\rm cm^2}$ is the Balmer-edge bound-free cross-section scale.  The cross-section is fixed by atomic physics; the spectrum constrains $\tau_{\rm B}$, and the thick-disk model restricts the allowed $N_{\rm H,B}$.  The best-fit $\tau_{\rm B}=15$ is naturally associated with a high-column, near-edge-on sight line.  Taking $N_{\rm H,B}\simeq N_{\rm H,edge}=2.5\times10^{25}{\rm cm^{-2}}$ from Equation~\ref{eq:td_nhedge} implies
\begin{equation}
\frac{N_2}{N_{\rm H}}\simeq
\frac{\tau_{\rm B}}{N_{\rm H,B}\sigma_{\rm B}}
=3.0\times10^{-6}.
\label{eq:td_n2frac}
\end{equation}
This small excited-state fraction is the consistency requirement imposed by the observed Balmer curvature.  If instead the vertical column were used, the same excited-state fraction would give $\tau_{\rm B}\simeq15N_{\rm H,\perp}/N_{\rm H,edge}\simeq8.7$, so the stronger fitted Balmer curvature supports the orientation picture in which RUBIES-EGS-55604 is viewed through the disk body or a high-covering disk atmosphere.  This is not a first-principles ionization calculation, but it is a useful consistency check linking the spectral fit to the column-density model.

\begin{table}
\centering
\small
\caption{Continuum-model parameters for RUBIES-EGS-55604.  ``Derived'' means fixed by the thick-disk/SIDM equations in Section~2 once the host halo, compact-disk mass, compact-disk speed, and $\sigma/m$ are specified.  ``Effective fit'' means constrained by the line-masked continuum because the parameter depends on unresolved radiative transfer, viewing angle, and ionization structure.}
\label{tab:55604_spectrum_params}
\begin{tabular}{p{0.20\textwidth}p{0.21\textwidth}p{0.25\textwidth}p{0.22\textwidth}}
\toprule
Parameter & Value & Role & Status \\
\midrule
$M_{\rm h}$ & $5.0\times10^{10}M_\odot$ & host halo scale & model input \\
$C_{\rm s}$ & $84.9{\rm km\,s^{-1}}$ & SIDM sound speed & derived \\
$M_{\rm gas,d}$ & $10^8M_\odot$ & compact inflow disk & simulation-motivated input \\
$R_{\rm d}$ & $14$ pc & disk size & derived from $GM_{\rm gas,d}/v_{\rm c,d}^2$ \\
$N_{\rm H,edge}$ & $2.5\times10^{25}{\rm cm^{-2}}$ & high-column sight line & derived \\
$\sigma/m$ & $1{\rm cm^2\,g^{-1}}$ & galaxy-scale SIDM cross-section & model input \\
$M_{\rm BH,SIDM}$ & $9.0\times10^7M_\odot$ & saturated SIDM-grown black hole & derived \\
$\lambda_{\rm eff}$ & 0.6 & buried accretion state & continuum constrained \\
$L_{\rm bol}$ & $6.8\times10^{45}{\rm erg\,s^{-1}}$ & AGN power & derived from $\lambda_{\rm eff}L_{\rm Edd}$ \\
$\alpha_{\rm red}$ & 4.3 & obscured AGN slope & effective fit \\
$f_{\rm red}$ & 0.0245 & emergent obscured-AGN fraction & effective fit \\
$T_{\rm gas}$ & $1.4\times10^4$ K & dense-gas color temperature & effective fit \\
$f_{\rm ph}$ & 0.0455 & obscured-photosphere luminosity fraction & effective fit \\
$C_{\rm B}$ & 0.70 & Balmer-layer covering factor & effective fit \\
$\tau_{\rm B}$ & 15 & Balmer-edge optical depth & effective fit, but model constrained\\
$N_2/N_{\rm H}$ & $3.0\times10^{-6}$ & required excited-H fraction & derived from $\tau_{\rm B}/N_{\rm H,edge}\sigma_{\rm B}$ \\
$\Delta\lambda_{\rm B}$ & $0.006\,\mu{\rm m}$ & Balmer-edge smoothing width & effective fit \\
$\mathrm{SFR}_{\rm eff}$ & $6.9M_\odot{\rm yr^{-1}}$ & leaked young stars & aperture nuisance, subdominant \\
$M_{\star,\rm old}$ & $7.6\times10^8M_\odot$ & old host light & aperture nuisance, subdominant \\
\bottomrule
\end{tabular}
\end{table}

The fitted effective values have simple physical interpretations.  The steep $\alpha_{\rm red}$ is the rest-optical tail of radiation that has been scattered, absorbed, and re-emitted through a high-column disk body.  The small $f_{\rm red}$ and $f_{\rm ph}$ keep the emergent continuum below the bolometric power available from the buried AGN, so no extra energy source is introduced.  The fitted $T_{\rm gas}$ is close to the characteristic temperature of dense photoionized gas and sets the blue color of the obscured photosphere.  The fitted $C_{\rm B}$ and $\tau_{\rm B}$ encode the fact that the Balmer-active layer covers much, but not all, of the continuum source; increasing them deepens the Balmer-limit curvature, while changing the edge width controls how sharp the break appears after velocity gradients, scattering, and instrumental resolution are folded together.  The stellar terms are allowed because the RUBIES aperture contains the compact galaxy, but their fitted amplitudes remain perturbative.

\begin{figure}
\centering
\includegraphics[width=0.95\textwidth]{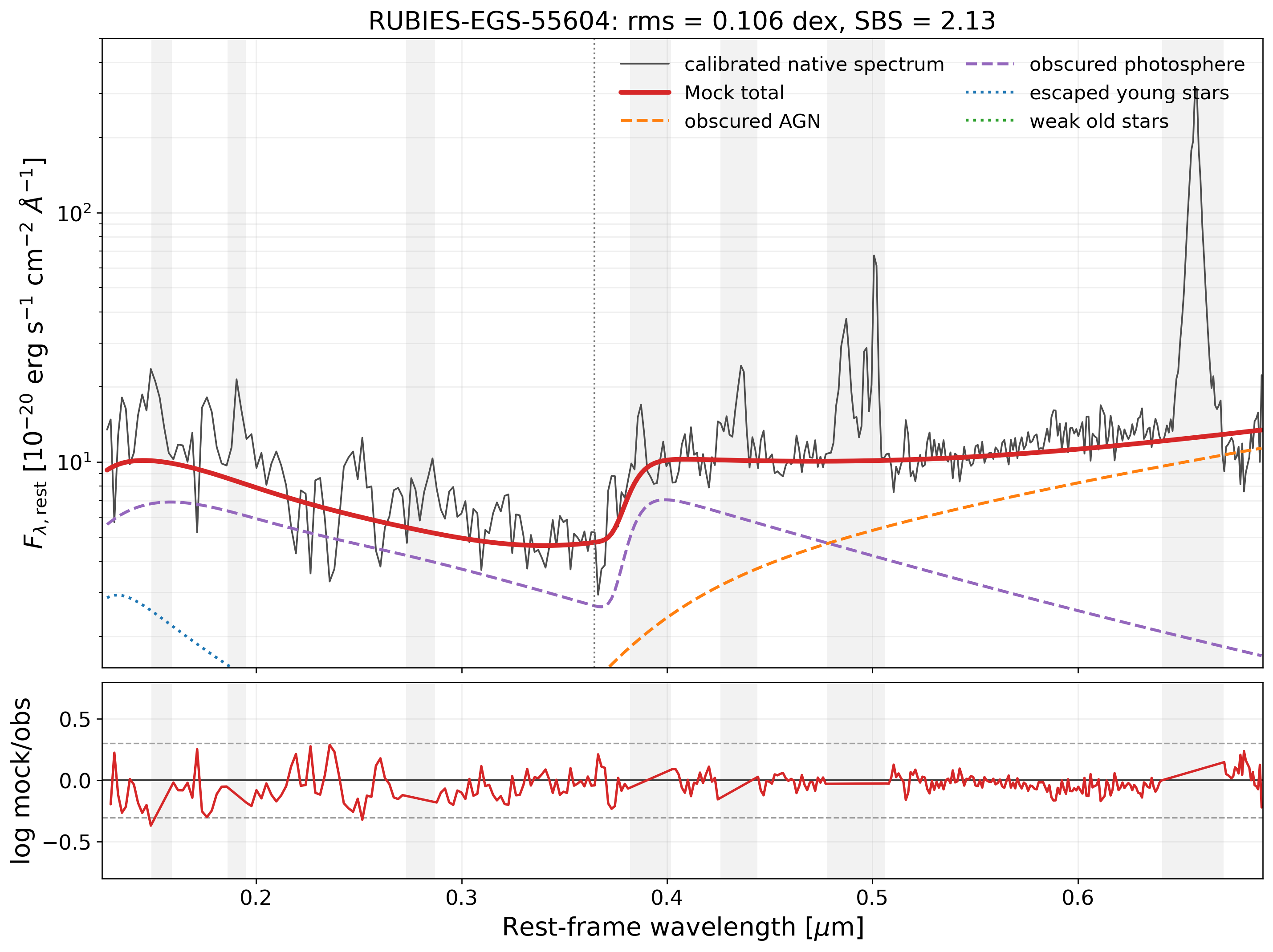}
\caption{{\bf{Mock spectrum versus the calibrated RUBIES-EGS-55604 spectrum. }} The red curve is the Mock total; dashed or dotted curves show the obscured AGN, smooth Balmer-transmission obscured photosphere, leaked young-star component, and weak old host component.  Grey regions are masked emission-line complexes.  The fit has rms residual $0.106$ dex outside the masked windows and spectral-break strength ${\rm SBS}=2.13$, matching the calibrated observed value.}
\label{fig:td_spectra}
\end{figure}

The plotted fluxes in Fig.~\ref{fig:td_spectra} are absolute predictions for $z=6.982$ and are not renormalized to the observed RUBIES-EGS-55604 spectrum.  To match the convention used in the RUBIES spectral figures, the model (red spectrum in Fig.~\ref{fig:td_spectra})  and data (black spectrum) are displayed as rest-frame-density $F_\lambda$ in units of $\mathrm{erg\,s^{-1}\,cm^{-2}}\,\text{\AA}^{-1}$; the underlying model normalization is computed in observed $F_\nu$, converted to observed $F_\lambda$ using $F_{\lambda,\rm obs}=F_{\nu,\rm obs}c/\lambda_{\rm obs}^2$, and then multiplied by $(1+z)$ to report the rest-frame-density convention.  Emission-line windows are masked when computing continuum residuals because reproducing line profiles requires photoionization and dynamical radiative-transfer modeling beyond the present continuum demonstration \citep{OsterbrockFerland2006,Ferland2017Cloudy}.  The masked windows include the UV high-ionization features, C~III], Mg~II, the H$\zeta$/H$\epsilon$/[Ne~III]/He~I complex, H$\gamma$/[O~III], H$\beta$/[O~III], and H$\alpha$/[N~II] complexes identified in the RUBIES spectra \citep{Wang2024RUBIESLRD}.

The fit uses the spectral-break-strength definition of \citet{Wang2024RUBIESLRD}, approximately the ratio of the continuum just redward and blueward of the Balmer limit.  The model gives ${\rm SBS}=2.13$, while the calibrated spectrum gives ${\rm SBS}=2.13$.  The stellar terms remain perturbative: the best fit uses a modest leaked young-star component and a weak old host contribution.  The V-shaped continuum is therefore produced primarily by the obscured AGN plus the smooth Balmer-transmission photosphere, not by an old stellar population.

\subsection{A Cooler LRD: RUBIES-UDS-40579}

\begin{figure}
\centering
\includegraphics[width=0.95\textwidth]{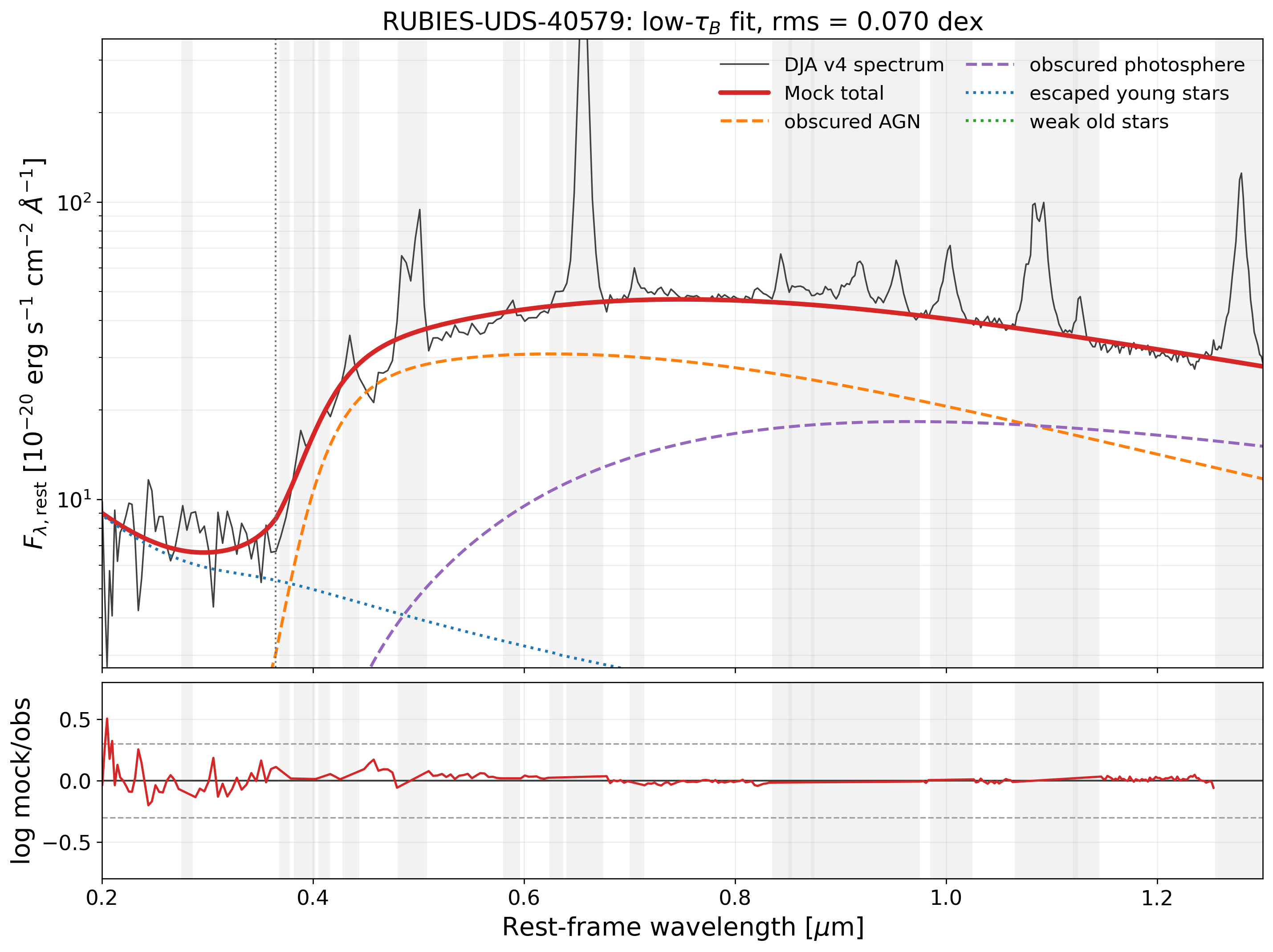}
\caption{{\bf{Mock spectrum compared with the observed spectrum of RUBIES-UDS-40579.}}  Analogous to Fig.~\ref{fig:td_spectra}, the red curve is the Mock total.  The orange curve is the obscured AGN continuum after passing through the same smooth Balmer-active disk atmosphere as the photosphere, and it contributes $58\%$ of the red continuum over $0.75$--$0.90\,\mu{\rm m}$.  The purple curve is the cooler obscured photosphere powered by absorbed AGN radiation.  The blue curve is a modest leaked young-star component.  Grey regions are masked emission-line complexes.}
\label{fig:td_40579}
\end{figure}

We next apply the same model to RUBIES-UDS-40579, one of the high-signal spectra in the de Graaff et al. LRD sample \citep{deGraaff2025LRDPopulation}. RUBIES-UDS-40579 represents a class of LRD that exhibit a red spectral bump and may have relatively low temperatures. This object has $z=3.114$ and a fitted modified-blackbody temperature $T_{\rm MBB}=3648$ K \citep{deGraaff2025LRDPopulation}.  We use the public DJA v4 PRISM spectrum directly, convert the observed-frame $F_\nu$ to observed-frame $F_\lambda$, and compare the model in the same flux-density convention, as shown in Fig.~\ref{fig:td_40579}.  When a rest-frame luminosity density is needed, we use
\begin{equation}
L_{\lambda,\rm rest}=4\pi D_L^2(1+z)F_{\lambda,\rm obs}.
\end{equation}
As for RUBIES-EGS-55604, emission-line windows are masked and only the continuum is fitted.

The best continuum solution has
\begin{equation}
M_{\rm h}=1.0\times10^{11}M_\odot,\quad
\sigma/m=1.0{\rm cm^2\,g^{-1}},\quad
\lambda_{\rm eff}=0.6,
\end{equation}
which gives $C_{\rm s}=77.1{\rm km\,s^{-1}}$, $M_{\rm BH,SIDM}=6.1\times10^7M_\odot$, $t_{\rm SIDM}=0.29$ Myr, and $L_{\rm bol}=4.6\times10^{45}{\rm erg\,s^{-1}}$.  
The compact disk parameters are $M_{\rm gas,d}=3.0\times10^7M_\odot$, $v_{\rm c,d}=100{\rm km\,s^{-1}}$, $R_{\rm d}=12.9$ pc, $N_{\rm H,edge}=5.1\times10^{24}{\rm cm^{-2}}$, and $\tau_T=3.4$.

For this source we use the AGN-dominated version of the same high-column smooth-Balmer model.  The red continuum is not required to be dominated by the thermalized photosphere.  Instead, an obscured AGN continuum provides most of the red-side continuum, while the obscured photosphere represents the fraction of absorbed power that escapes as a cooler compact thermal component.  Both components propagate through the same outer Balmer-active disk atmosphere described by Equation~\ref{eq:smooth_Balmer_trans}.  
We write the AGN continuum as
\begin{equation}
F_\nu^{\rm obscured\ AGN}\propto
F_\nu^{\rm red}(\alpha_{\rm red})
\left[1+\left(\frac{\lambda}{\lambda_{\rm turn}}\right)^{s_{\rm turn}}\right]^{-1}
{\cal T}_{\rm B}(\lambda),
\label{eq:td_agn_turnover}
\end{equation}
where $\lambda_{\rm turn}$ and $s_{\rm turn}$ describe the red-side turnover caused by self-absorption, diffusion, or reprocessing in the optically thick disk body.  This is still an effective radiative-transfer template; it is not a new dynamical component.  In RUBIES-EGS-55604 the turnover is effectively unconstrained because any acceptable turnover lies far to the red of the fitted wavelength range.  In RUBIES-UDS-40579 the available spectrum extends farther into the rest optical and therefore weakly constrains the turnover.

The best obscured-AGN continuum solution has
\begin{equation}
\alpha_{\rm red}=3.0,\quad
\lambda_{\rm turn}=0.82\,\mu{\rm m},\quad
s_{\rm turn}=4.0,
\end{equation}
for the obscured AGN and
\begin{equation}
T_{\rm gas}=3000{\rm K},\quad
\beta_{\rm ph}=0,
\end{equation}
for the obscured photosphere.  The fitted luminosity fractions are
\begin{equation}
f_{\rm AGN}=0.055,
\qquad
f_{\rm ph}=0.051,
\end{equation}
where $f_{\rm AGN}$ is the emergent obscured-AGN fraction and $f_{\rm ph}$ is the emergent obscured-photosphere fraction.  The smooth Balmer layer has $\tau_{\rm B}=3.0$, $C_{\rm B}=0.95$, and $\Delta\lambda_{\rm B}=0.025\,\mu{\rm m}$.  For the 40579 disk $H/R_{\rm d}\simeq1$, so the vertical and disk-body columns are both $N_{\rm H,B}\sim5.1\times10^{24}{\rm cm^{-2}}$ and the corresponding Thomson depth is $\tau_{\rm T,edge}\simeq3.4$.  The fitted Balmer depth is therefore of the same order as the column-density estimate and requires only $N_2/N_{\rm H}\simeq2.9\times10^{-6}$.  The blue side is supplied by a modest leaked young-star component with effective escaped ${\rm SFR}=6.8M_\odot{\rm yr^{-1}}$, age $50$ Myr, and $E(B-V)_\star=0.2$.  The old host component is again included from the seed-host prior, $M_{\star,\rm old}=3.3\times10^7M_\odot$, but is optically weak.

The line-masked continuum rms is $0.070$ dex.  The red-side AGN contribution is $58\%$ over $0.75$--$0.90\,\mu{\rm m}$ and $58\%$ over the masked continuum as a whole, so the fit is genuinely AGN dominated in the red continuum.  
This experiment shows that RUBIES-UDS-40579 does not require a photosphere-dominated red continuum; an Eddington-subcritical buried AGN with $\lambda_{\rm eff}=0.6$ can dominate the red side when it passes through the outer Balmer shell.

The fitted $T_{\rm gas}=3000$ K is consistent with a compact dense-gas color temperature.  Since $f_{\rm ph}=0.051$ is only the observed escaping photosphere fraction, the obscuring disk must thermalize substantially more power than this in the unseen channels.  Thus the cool photosphere temperature is set by the absorbed AGN heating and the unresolved radiative-transfer state of the disk atmosphere, not by equating the observed photosphere luminosity alone to the total heating power.  The contrast with the $T_{\rm gas}=1.4\times10^4$ K 55604 solution is therefore allowed within the same model: 55604 is a hotter, more photoionized or shallower escaping layer, and possibly corresponds to an earlier stage of LRD, whereas 40579 is a cooler, more shielded, redder effective layer of the same compact thick disk, which may correspond to a later stage of LRD.

\begin{table}
\centering
\small
\caption{Best-fit physical and effective continuum parameters for the two LRD spectra fitted in this work.  The first block is tied to the thick-disk/SIDM model; the second block contains effective continuum-transfer parameters fitted to the line-masked continuum.  RUBIES-UDS-40579 is shown with the obscured-AGN red-continuum solution.}
\label{tab:rubies_fits}
\begin{tabular}{lcc}
\toprule
Parameter & RUBIES-EGS-55604 & RUBIES-UDS-40579 \\
\midrule
$z$ & 6.982 & 3.113558 \\
$M_{\rm h}$ & $5.0\times10^{10}M_\odot$ & $1.0\times10^{11}M_\odot$ \\
$M_{\rm gas,d}$ & $10^8M_\odot$ & $3.0\times10^7M_\odot$ \\
$v_{\rm c,d}$ & $175{\rm km\,s^{-1}}$ & $100{\rm km\,s^{-1}}$ \\
$R_{\rm d}$ & 14 pc & 12.9 pc \\
$N_{\rm H,edge}$ & $2.5\times10^{25}{\rm cm^{-2}}$ & $5.1\times10^{24}{\rm cm^{-2}}$ \\
$\sigma/m$ & $1.0{\rm cm^2\,g^{-1}}$ & $1.0{\rm cm^2\,g^{-1}}$ \\
$C_{\rm s}$ & $84.9{\rm km\,s^{-1}}$ & $77.1{\rm km\,s^{-1}}$ \\
$M_{\rm BH,SIDM}$ & $9.0\times10^7M_\odot$ & $6.1\times10^7M_\odot$ \\
$t_{\rm SIDM}$ & 0.32 Myr & 0.29 Myr \\
$\lambda_{\rm eff}$ & 0.6 & 0.6 \\
$L_{\rm bol}$ & $6.8\times10^{45}{\rm erg\,s^{-1}}$ & $4.6\times10^{45}{\rm erg\,s^{-1}}$ \\
\midrule
$\alpha_{\rm red}$ & 4.3 & 3.0 \\
$\lambda_{\rm turn}$ & -- & $0.82\,\mu{\rm m}$ \\
$s_{\rm turn}$ & -- & 4.0 \\
$T_{\rm gas}$ & $1.4\times10^4$ K & $3000$ K \\
$\beta_{\rm ph}$ & 0 & 0 \\
$f_{\rm red/AGN}$ & 0.0245 & 0.055 \\
$f_{\rm ph}$ & 0.0455 & 0.051 \\
$C_{\rm B}$ & 0.70 & 0.95 \\
$\tau_{\rm B}$ & 15 & 3.0 \\
$N_2/N_{\rm H}$ required & $3.0\times10^{-6}$ & $2.9\times10^{-6}$ \\
$\Delta\lambda_{\rm B}$ & $0.006\,\mu{\rm m}$ & $0.025\,\mu{\rm m}$ \\
$\mathrm{SFR}_{\rm eff}$ & $6.9M_\odot{\rm yr^{-1}}$ & $6.8M_\odot{\rm yr^{-1}}$ \\
$f_{\rm esc,young}$ & 0.2 & 1.0 \\
$E(B-V)_\star$ & 0 & 0.2 \\
$M_{\star,\rm old}$ & $7.6\times10^8M_\odot$ & $3.3\times10^7M_\odot$ \\
\midrule
AGN fraction, red continuum & 0.72 & 0.58 \\
Continuum rms & 0.106 dex & 0.070 dex \\
\bottomrule
\end{tabular}
\end{table}

\section{Discussion}

\subsection{Comparison with Other Proposed LRD Channels}

The model developed here belongs to a broader family of explanations in which LRDs are short-lived phases rather than a new long-lived galaxy class.  The most useful way to compare these ideas is not only by asking whether they can make a red compact spectrum, but by asking what they require of the host halo population.  Recent clustering and assembly-bias calculations show that several proposed LRD channels select very special halos and therefore predict strong or distinctive clustering: direct-collapse black holes, SIDM core collapse, low-spin compact-galaxy models, and primordial-black-hole-related channels can all carry a large host-halo or assembly-bias burden \citep{CarranzaEscudero2025LRDclustering,Wang2026LRDAssemblyBias,Zhang2026LRDhalo}.  This is a serious discriminant because current LRD environment measurements do not yet require every LRD to live in the rarest, most biased peaks.  Our model is designed to avoid this problem.  It does not identify LRDs with a globally extreme halo class.  The halo provides the SIDM reservoir and velocity scale, while the LRD switch is local: a gas-rich compaction event builds a compact nuclear thick disk, compresses the central SIDM, and then hides the gas-powered AGN along high-column sight lines.

This point is sharpest for the SIDM gravothermal core-collapse scenario of \citet{Jiang2026SIDMLRD}.  That model elegantly produces massive black holes without luminous super-Eddington growth, but the collapse time is extremely sensitive to halo concentration, formation time, and assembly history.  The same sensitivity that helps make early black holes also tends to select unusually early or concentrated halos, producing a strong environmental prediction \citep{Wang2026LRDAssemblyBias}.  It also leaves a redshift question: if halo age alone controls the visible event, lower-redshift halos have had more time to collapse, so the disappearance of LRDs toward low redshift is not automatic.  In our picture SIDM still supplies the rapid dark mass growth, but the visible LRD phase is gated by high-redshift gas compaction and column density.  Low-redshift halos may contain SIDM cores or black holes, but they usually lack the compact, gas-rich, Compton-thick nuclear disk needed to appear as an LRD.

Direct-collapse black-hole and quasi-star scenarios face a complementary problem \citep{Mayer2014DirectCollapse,Mayer2023CosmoDCBH,Jeon2025DCBHLRD,BegelmanDexter2025QuasiStars,PacucciFerraraKocevski2026DCBH}.  They naturally produce buried accretion and dense gas, but the required initial conditions are severe: rapid inflow, suppressed fragmentation, special thermodynamic histories, and often a restricted halo population.  If most LRDs were direct-collapse remnants, their abundance and clustering would trace the rarity of those conditions.  Our model keeps the robust part of the direct-collapse simulations--the formation of a compact turbulent nuclear disk--but does not require the disk to form a black-hole seed already close to the observed LRD black-hole mass.  It still requires a pre-SIDM seed massive enough for the compressed SIDM density to satisfy the ignition condition in Equation~\ref{eq:td_ignition}; in the fiducial model this is of order $10^5M_\odot$, while denser nuclear disks can lower the requirement toward $10^4M_\odot$ or even Pop~III-remnant scales.  This seed requirement is therefore comparatively mild and can plausibly be met in ordinary low-mass high-redshift galaxies or halos, rather than only in pristine direct-collapse sites.  The large black-hole mass jump then occurs through SIDM accretion once the disk compresses the central dark matter, and the saturation mass is set by Equation~\ref{eq:td_msat}.  Thus the dense gas geometry is inherited from the simulations, while the need for a seed already comparable to the observed LRD black-hole mass is avoided.

Pure obscured-AGN or dusty-flow models solve the spectral problem with familiar physics but put too much work on gas accretion alone \citep{Greene2024LRD,Li2024DustyFlows,PacucciNarayan2024XrayWeak}.  A particularly clear example is the super-Eddington unification model of \citet{MadauMaiolino2026LRDLBD}, in which LRDs are obscured little blue dots viewed through a dusty equatorial screen around a thick accretion flow, while polar sight lines reveal the blue phase.  This picture usefully captures orientation dependence and the connection between red and blue compact AGN.  Its main burden is different from ours: the overmassive black hole must still be built by luminous gas accretion, usually in a super-Eddington state, and the model does not by itself explain why the phase is tied to a finite high-redshift gas-compaction event.  More generally, in pure gas-accretion models the same gas must grow an overmassive black hole, power the observed continuum, remain X-ray weak, and avoid rapidly clearing itself by feedback.  This often pushes the model toward high duty cycles, super-Eddington growth, or carefully tuned covering factors.  Our model separates the tasks.  SIDM accretion makes the black hole overmassive while radiating little after a finite ignition threshold is crossed; subsequent baryonic accretion can be sub-Eddington or near-Eddington; and the nuclear thick disk supplies the Compton-thick column that suppresses escaping X-rays.  The X-ray weakness is therefore not an independent patch, but a consequence of the same gas structure that triggered the event.

Compact stellar, low-spin compact-galaxy, and forming-globular-cluster models emphasize real and important facts: LRDs are small, dense, and can contain stellar light \citep{Williams2024CompactLRD,Baggen2024LRDDensities,PacucciLoeb2025LowSpin,Chisholm2026GlobularLRD}.  Their weakness is that stellar compactness alone does not naturally explain broad permitted lines, dynamically or virially inferred overmassive black holes, weak X-rays, and a short high-redshift duty cycle.  Low-spin models also risk becoming another special-halo selection, with corresponding clustering predictions \citep{Wang2026LRDAssemblyBias}.  In our fits the young and old stellar components are allowed, because the spectroscopic aperture includes the compact host, but they remain perturbative.  Stars trace the inflow; they do not have to power the LRD.

Dense-gas or black-hole-star interpretations correctly identify an essential radiative-transfer clue: the Balmer-limit curvature and absorption can arise from dense gas rather than an old stellar population \citep{InayoshiMaiolino2024DenseGas,deGraaff2025TheCliff}.  We adopt this lesson, but add a dynamical trigger.  In our model the Balmer-active layer is the atmosphere or wind base of the same Compton-thick nuclear disk that compressed the SIDM.  The fitted Balmer optical depths are checked against the disk columns, and the required excited-state fractions are small.  Dense gas is therefore not just a spectral fitting component; it is the physical mediator between baryonic inflow, SIDM compression, black-hole growth, and obscuration.

Finally, primordial-black-hole, dark-star, and other exotic channels can make early compact massive objects \citep{Carr2016PBH,ZhangFengAn2026PBHClusters,Ilie2026DarkStars}, but they still must explain why the observed objects have LRD spectra, why the phase is X-ray weak, why it fades at low redshift, and why its clustering is not obviously extreme.  Our mechanism is less radical in its initial conditions and more restrictive in its observable trigger.  It predicts that LRDs should correlate most strongly with local compaction, disturbed gas, compact high-column nuclei, and orientation, rather than only with rare halo labels.  This is the central advantage of the model: it can produce overmassive black holes without demanding that every LRD occupy a globally exceptional halo, and it connects that mass growth to the same gas structure that makes the source red.

\subsection{Observational Predictions}

The model makes several falsifiable predictions.  First, LRDs should be orientation dependent.  High-column sight lines through the disk body, inner rim, or turbulent atmosphere appear red and X-ray weak, whereas polar sight lines should reveal bluer compact AGN with weaker Balmer curvature.  Second, the strongest LRD signatures should correlate with compact gas, disturbed morphology, close companions, high attenuation, or other signs of recent inflow rather than only with large-scale halo overdensity.  Third, the LRD phase should be short.  Once the high-covering column declines, the polar channel widens, or the Balmer-active layer disappears, the descendant should be a less-obscured compact AGN, a blue compact broad-line AGN, or a compact post-starburst system with an overmassive black hole.  Fourth, cooler LRDs such as RUBIES-UDS-40579 may represent more shielded or later radiative-transfer states of the same thick disk, while hotter V-shaped sources such as RUBIES-EGS-55604 may trace a more strongly photoionized or higher-column state.

This framework also explains why low-redshift analogues are uncommon.  Later dwarf and small-galaxy mergers are less gas rich, more feedback regulated, and less able to maintain a compact Compton-thick nuclear disk.  Moreover, once the first large fractional SIDM-fed black-hole growth event has occurred, later inflows change the black-hole mass by a smaller factor and are less able to produce the same potential-transformation phase.  In more massive halos the velocity-dependent SIDM cross-section can be smaller, further suppressing efficient SIDM accretion.  Thus the absence of abundant low-redshift LRDs follows from gas supply, orientation, duty cycle, and the one-time nature of the first large SIDM-fed mass jump.

\subsection{Limitations and Next Steps}

The present model is deliberately semi-analytic.  The continuum parameters $\alpha_{\rm red}$, $f_{\rm red}$, $f_{\rm ph}$, $T_{\rm gas}$, $C_{\rm B}$, $\tau_{\rm B}$, and the Balmer-edge width obey energy and column-density constraints, but they are effective radiative-transfer parameters.  A full radiation-hydrodynamic and photoionization calculation is required to predict them from first principles \citep{Ferland2017Cloudy}.  The stellar leakage and old-host terms are aperture-level nuisance components rather than central ingredients of the model.  The SIDM compression calculation assumes an approximately isothermal SIDM response and parameterizes the disk potential with $\eta_\Phi$.  The decisive next step is a coupled simulation of a compact baryonic disk embedded in an SIDM halo with a growing central black hole, followed by radiative-transfer modeling of spectra, emission lines, and X-rays.

\section{Conclusions}

We have proposed a gas-triggered SIDM-accretion model for Little Red Dots.  The central idea is that a compact nuclear thick disk can do two things at once: compress the central SIDM enough to ignite rapid dark accretion onto a pre-existing black hole, and provide the high-column gas that later reprocesses AGN radiation into an LRD spectrum.  For a fiducial RUBIES-EGS-55604-like halo, the model gives $M_{\rm BH}\simeq9\times10^7M_\odot$ in $0.32$ Myr for $\sigma/m=1\,{\rm cm^2\,g^{-1}}$, while the subsequent gas-powered luminosity remains sub-Eddington.

The model naturally links several observed LRD properties.  The black hole can become overmassive without requiring hyper-Eddington radiative growth.  The compact disk provides Compton-thick columns, explaining weak escaping X-rays and strong reprocessing.  A smooth Balmer-active disk atmosphere produces the Balmer-limit curvature.  Orientation determines whether the same event appears as a red LRD or a bluer compact AGN.  The LRD phase ends when the high-covering column is consumed, expelled, diluted, or no longer replenished, giving a natural lifetime of tens to hundreds of Myr.

Using the same thick-disk reprocessing framework, we fit two representative spectra: RUBIES-EGS-55604, a hotter V-shaped LRD, and RUBIES-UDS-40579, a cooler red-bump LRD.  The required Balmer optical depths are consistent with the columns predicted by the disk model and imply excited-hydrogen fractions of order $10^{-6}$.  These fits are not a full radiative-transfer solution, but they demonstrate that the proposed physical picture can reproduce the observed continuum diversity with plausible parameters.

The most important observational tests are environmental and temporal.  If our model is correct, LRDs should preferentially trace compact inflow events, close companions, disturbed morphologies, high covering columns, and orientation-dependent obscuration, rather than only the rarest high-concentration halos.  Future JWST spectroscopy, X-ray constraints, MIRI photometry, resolved morphology, and clustering measurements can therefore distinguish gas-triggered SIDM accretion from direct collapse, gravothermal core collapse, compact stellar systems, and other proposed channels.

\begin{acknowledgments}
YR acknowledges support from the CAS Pioneer Hundred Talents Program (Category B), the NSFC grants 12522302, 12673017 and 12273037, and the USTC Research Funds of the Double First-Class Initiative.
\end{acknowledgments}

\bibliography{references}

\end{document}